\documentclass{svjour2}                    % onecolumn
\smartqed  % flush right qed marks, e.g. at end of proof
\usepackage{graphicx}
\usepackage{latexsym}
\begin{document}

\title{Correlation Effects on the Charge Radii of Exotic Nuclei}  



\author{TOMASELLI \and K{\"U}HL  \and URSESCU  \and FRITZSCHE}


\institute{M. TOMASELLI
           \at Physics Department, TUD Darmstadt, D-64289-Darmstadt, Germany \\
               Gesellschaft f{\"u}r Schwerionenforschung (GSI), 
	       D-64291 Darmstadt, Germany \\
               \email{m.tomaselli@gsi.de}      
           \and
           T. K{\"U}HL  
	   \at Gesellschaft f{\"u}r Schwerionenforschung (GSI), 
	       D-64291 Darmstadt, Germany \\
           \and 
	   D. URSESCU 
	   \at Gesellschaft f{\"u}r Schwerionenforschung (GSI), 
	       D-64291 Darmstadt, Germany \\
           \and 
	   S. FRITZSCHE 
	   \at Institut f\"u{}r Physik, Kassel University, D-34132 Kassel, 
	       Germany }            

\date{Received: date / Accepted: date}

\maketitle

\begin{abstract}
The structures and distributions of light nuclei are investigated within 
a microscopic correlation model. Two particle correlations are responsible 
for the scattering of model particles either to low momentum-
or to high momentum-states. The low momentum states form the model space 
while the high momentum states are used to calculate the G-matrix. 
The three and higher order particle correlations do not play a role in
the latter calculation especially if the correlations induced
by the scattering operator are of sufficient short range. They modify however, 
via the long tail of the nuclear potential, the Slater determinant of the 
(A) particles by generating excited Slater's determinants.

\end{abstract}

\newpage
\section{Introduction}
\label{intro}

Correlation effects in nuclei have first been introduced in nuclei by 
Villars~\cite{vil63}, who proposed the unitary-model operator (UMO) to 
construct effective operators. The method was implemented by 
Shakin~\cite{sha66,sha68} for the calculation of the G-matrix from hard-core 
interactions.

The UMO is based on the separation of the two body potential in a short and 
a long components. Within this separation the effective n-body Hamiltonian 
contains only the long component. The short-range component is considered 
up to the two body correlation and produces no energy shift in the pair 
state.

Non perturbative approximations of the UMO have been recently applied to 
even nuclei in Ref.~[4,5] which is treated here in more detail. 
The basics formulas of the Boson Dynamic Correlation Model (BDCM) presented 
in the above quoted paper have been obtained by solving the n-body problem 
in terms of the long range component of the two-body force. 
This component has the effect of generating a new correlated model space 
(effective space) which departs from the originally adopted one (shell model).
The amplitudes of the model wave functions are calculated in terms of
non linear equation of motions (EoM).

By linearizing the systems of commutator equations, which characterize the 
EoM, we derive the eigenvalue equations for our model space.
 Within this correlated formalism we generate a model that includes not 
only the ladder diagrams of Ref.~[6] but also the folded diagrams of 
Kuo~\cite{kuo01}.

The n-body matrix elements which define the eigenvalue equations
are calculated exactly via the Cluster Factorization Theory 
(CFT)~\cite{tom06}.

In this paper the BDCM model is applied to calculate the influence of the 
correlations on the charge distributions of the lithium isotopes and on the 
charge distribution of $^6$He.

The value obtained for the charge radius of the correlated $^6$He is slightly
bigger than the radius calculated in other theories~[9-13]
and that derived within the isotopic-shift IS theory~\cite{wan04}. A charge 
radius which agrees  with the radii calculated in the Refs.~[9-13] and 
those calculated in the cluster models of Refs.~[15-17]  is on the other hand 
obtained by considering only two protons in the $s_{\frac{1}{2}}$.
This non correlated radius agrees also with the radius derived at Argonne
within the IS theory~\cite{wan04}. Correlations have therefore the property 
to increase the charge radius of $^6$He as observed for the isotopes of 
Lithium.

The calculations performed in Ref.~[18] for the charge radii of the lithium 
isotopes, although in good agreement with those measured at 
GSI-TRIUMF~[19-21] and analyzed with the help of Ref.~[22], are always 
slightly larger than those measured. For the stable isotope $^6$Li the 
calculated radius agrees with the value obtained with the electron scattering 
experiments of  Ref.~[23]. However, the charge radii calculated in the IS 
theory could also depend on the nuclear correlations. The calculations 
performed in Ref.~[24] for the field shift (FS) of $^7$Li show that the 
departure from a point nuclear approximation is a rather large effect.   
Additionally the higher order cross term contributions of Ref.~[25] 
need to be considered.

A direct comparison between the calculated and the measured charge radii 
should therefore be performed after an accurate analysis of these two 
correcting factors.

\section{Theory of Two Correlated Particles}
\label{sec:1}
In order to describe the structures and the distributions of nuclei
we start from the following Hamiltonian:
\begin{equation}\label{equ.1}
H= \sum_{\alpha\beta}
 \langle\alpha|t|\beta\rangle \, a^{\dagger}_{\alpha}a_{\beta}
   \:+\: \sum_{\alpha\beta\gamma\delta}
   \langle\Phi_{\alpha\beta}| v_{12}|\Phi_{\gamma\delta}\rangle \,
   a^{\dagger}_{\alpha}a^{\dagger}_{\beta} a_{\delta}a_{\gamma}  
\end{equation}
were $v_{12}$ is the singular nucleon-nucleon two body potential.
Since the matrix elements $|\alpha\beta\rangle$ are uncorrelated the matrix elements of 
$v_{12}$ are infinite. This problem can be avoided by taking matrix elements
of the Hamiltonian between correlated states.
In this paper the effect of correlation is introduced via the $e^{iS}$ method.
In dealing with very short range correlations only the $S_2$ part of the correlation
 operator need to be considered.  

Following Ref.~[2,3] we therefore calculate an ``effective Hamiltonian'' by using only the $S_2$ 
correlation operator obtaining:
\begin{equation}
\label{eq.1}
\begin{array}{l}
H_{eff}=e^{-iS_2}He^{iS_2}=\sum_{\alpha\beta}\langle\alpha|t|\beta\rangle a^{\dagger}_{\alpha}
a_{\beta}+
\sum_{\alpha\beta\gamma\delta}\langle\Psi_{\alpha\beta}|v^l_{12}|\Psi_{\gamma\delta}\rangle 
a^{\dagger}_{\alpha}a^{\dagger}_{\beta} a_{\delta}a_{\gamma}\\
=\sum_{\alpha\beta}\langle\alpha|t|\beta\rangle a^{\dagger}_{\alpha}a_{\beta}+
\sum_{\alpha\beta\gamma\delta}\langle\Psi_{\alpha\beta}|v|\Psi_{\gamma\delta}\rangle a^{\dagger}_{\alpha}a^{\dagger}_{\beta} a_{\delta}a_{\gamma}
\end{array}
\end{equation}
where $v_{12}^l$ is the long component of the two body interaction (note that the
$v^l_{12}$ is in the following equations simply denoted as v).
 The $\Psi_{\alpha\beta}$ is the two particle correlated wave function:
\begin{equation}\label{eq.1a}
\Psi_{\alpha\beta}=e^{iS_2}\Phi_{\alpha\beta}
\end{equation}

In dealing with complex nuclei however the ($S_i,~i=3\cdots n$)
correlations should also be considered.

The evaluation of these diagrams is, due to the 
exponentially increasing number of terms, difficult in a perturbation theory.

We note however that one way to overcome this problem is to work with 
$e^{i(S_1+S_2+S_3+\cdots+S_n)}$ operator on the Slater's determinant
 by keeping the n-body Hamiltonian unvaried.

After having performed the diagonalization of the n-body Hamilton's operator we
can calculate the form of the effective Hamiltonian which, by now, includes 
correlation operators of complex order.

Using Eq.\ (\ref{equ.1}), we can compute the commutator of the Hamiltonian with
the  operator $(a^{\dagger}_{j_1}a^{\dagger}_{j_2})^J$ that creates a valence particle pair.
By performing this calculation we shall retain the linear 
 and the non-linear terms which are formed
by coupling the shell model states with the particle-hole excitations of the core. 

At this point in order to obtain a complete system of equations,
we also have to calculate the
commutator of the Hamiltonian (\ref{equ.1}) with the non linear terms.
 With this calculation we introduce in the model space 
states which are formed by coupling the valence states   
 with the two particle-hole excitations of the core. 

The successive model equations are then formed by calculating the commutator
with the operator obtained in the previous step.

The set of commutator equations above is suitable to be solved by means 
of a perturbation expansion.
The perturbative solution of the system of commutator equations is however not
easily obtainable due to the high number of diagrams one needs to calculate. 

A non perturbative solution of the system of commutator equation
can be obtained within the linearization method, which consists by applying 
the Wick's theorem to the ((n+2)p-2h) terms and by neglecting the normal order terms.
 
This approximation is motivated by the consideration that
 the low lying spectra of nuclei the ((n+2)p-2h) terms are lying at 
much higher energy than that of the ((n+1)p-1h) states. 

The linearized system of the commutator equations is then solved exactly 
in terms of the CFT which calculates the n-body matrix elements in an 
expedite and exact way.

In the following we give the basic formula of
 the method for a (nuclear) system with
two valence particles. In second quantization, the two 
particle states are defined by:
\begin{equation}\label{e1}
   \Phi_{2p}\longrightarrow \Phi_{j_1j_2}^J=A^{\dagger}_1(\alpha_1J)|0\rangle
   = [a^{\dagger}_{j_1}a^{\dagger}_{j_2}]^J_M|0\rangle,
\end{equation}
where, for the sake of simplicity, we have omitted the isospin quantum numbers
and where
\begin{equation}\label{e3}
   \begin{array}{c}
   \alpha_1\leftrightarrow j_1j_2
   \end{array}
\end{equation}
has been introduced to ensure a compact index notation of the angular momenta 
of the two particles. In this notation, the operator product 
$a^{\dagger}_{j_1}a^{\dagger}_{j_2}$ just creates two \textit{coupled} 
particles of single particle $j_1$ and $j_2$ \textit{coupled} to the final J quantum
number. 
  
To derive the effect of the correlation on the two valence particles, we have,
at this stage, to evaluate the next commutator
\begin{eqnarray}\label{e2}
   &  & \hspace*{-1.0cm}
   [H,A^{\dagger}_1(\alpha_1J)]|0\rangle 
   \nonumber \\
   & = &
   [(\sum_{\alpha}\epsilon_{\alpha}a^{\dagger}_{\alpha}a_{\alpha}+\frac{1}{2}
   \sum_{\alpha \beta \gamma \delta} 
   \langle\alpha\beta|v(r)|\gamma\delta\rangle 
   a^{\dagger}_{\alpha} a^{\dagger}_{\beta}
   a_{\delta}a_{\gamma}),(a^{\dagger}_{j_1}a^{\dagger}_{j_2})^J]|0\rangle 
\end{eqnarray}
which, after some operator algebra becomes 
\begin{eqnarray}\label{e4}
   &  & \hspace*{-1.0cm}
   [H,A^{\dagger}_1(\alpha_1J)]|0\rangle
   \nonumber \\
   & = & \sum_{\beta_{1}} \Omega(2p|2p') A^{\dagger}_1(\beta_1J)]|0\rangle
   + \sum_{\beta_2J'_1J'_2} \Omega(2p|3p1h) 
   A^{\dagger}_2(\beta_2J'_1J'_2J)]|0 \rangle.
\end{eqnarray}

\vspace*{2.0cm}

In Eq.~(\ref{e4}) the $A^{\dagger}_1(\beta_1J)$ operators are those 
of Eq.~(\ref{e1}) and the  $A^{\dagger}_2(\beta_2J'_1J'_2J)$ are defined 
below:
\begin{equation}\label{e5}
   \begin{array}{l}
   \Phi_{3p1h}\longrightarrow \Phi_{j'_1j'_2j'_3j'_4}^{J'_1J'_2J}
   =A^{\dagger}_2(\beta_2J'_1J'_2J)|0\rangle
   = ((a_{j'_1}^{\dagger}a_{j'_2}^{\dagger})^{J'_1}
      (a_{j'_3}^{\dagger}a_{j'_4})^{J'_2})^J|0\rangle.
   \end{array}
\end{equation}
In Eq.~(\ref{e5}) we have used the additional convention:
\begin{equation}\label{e6}
   \begin{array}{c}
   \beta_2\longrightarrow j'_1j'_2j'_3j'_4
   \end{array}
\end{equation}
and we have associated:
\begin{equation}\label{e7}
   \begin{array}{cc}
   J'_1~$to the coupling of$~j'_1j'_2\\
   J'_2~$to the coupling of$~j'_3j'_4
   \end{array}
\end{equation}
Having extended the commutator as in Eq.~(\ref{e4}), we have also to calculate 
the commutator equation for the $A^{\dagger}_2(\alpha_2J_1J_2J)$ operators 
as given below: 
\begin{equation}\label{e8}
   \begin{array}{l}
   \displaystyle{
   [H,A^{\dagger}_2(\alpha_2J_1J_2J)]|0\rangle}  \\
   \displaystyle{
   = \sum_{\beta_{2}J'_1J'_2} \Omega(3p1h|3p'1h') 
     A^{\dagger}_2(\beta_2J'_1J'_2J)|0\rangle
   + \sum_{\beta_{3}J'_1J'_2J'_3} \Omega (3p1h|4p2h) 
     A^{\dagger}_3(\beta_3J'_1J'_2J'_3J)|0\rangle},
\end{array}
\end{equation}
where we have introduced the (4p-2h) wave functions defined below:
\begin{equation}\label{e9}
   \begin{array}{l}
   \Phi_{4p2h} \longrightarrow \Phi_{j'_1j'_2j'_3j'_4j'_5j'_6}^{J'_1J'_2J'_{12}J'_3J}
   =A^{\dagger}_3(\beta_3J'_1J'_2J'_3J)|0\rangle \\
   = (((a_{j'_1}^{\dagger}a_{j'_2}^{\dagger})^{J'_1}
   (a_{j'_3}^{\dagger}a_{j'_4})^{J'_2})^{J'_{12}}(a_{j'_5}^{\dagger}
    a_{j'_6})^{J'_3})^J|0\rangle,
\end{array}
\end{equation}
and where we have consistently extended the definition given 
in~(\ref{e3},\ref{e7}):
\begin{equation}\label{e10}
   \begin{array}{c}
   \beta_3\longrightarrow j'_1j'_2j'_3j'_4j'_5j'_6
   \end{array}
\end{equation}
with:
\begin{equation}\label{e11}
   \begin{array}{cc}
   J'_1~$associated to the coupling of$~j'_1j'_2\\
   J'_2~$associated to the coupling of$~j'_3j'_4\\
   J'_3~$associated to the coupling of$~j'_5j'_6
   \end{array}
\end{equation}
In the definition of $A^{\dagger}_3(\beta_3J'_1J'_2J'_3J)$ the coupling 
of $J'_1$ to $J'_2$ to $J'_{12}$ has been discarded from the notation.
In Eqs.~(\ref{e4},\ref{e8}) the $\Omega$'s are the matrix elements of the 
Hamilton's operator in the model wave functions. The next step would then be  
the computation of the commutator of the Hamiltonian with the 
$A^{\dagger}_3(\beta_3J'_1J'_2J'_3J)$ operators. Here we linearize these 
contributions by considering that in the study of the low energy spectrum 
and in the calculation of ground-state correlated distributions the 
$A^{\dagger}_3(\beta_3J'_1J'_2J'_3J)$ terms have a small contribution.
The linearization is performed by applying to the (4p2h) terms:
\begin{equation}\label{e12}
   \sum_{\alpha \beta \gamma \delta} 
   \langle\alpha\beta|v(r)|\gamma\delta\rangle a^{\dagger}_{\alpha} 
   a^{\dagger}_{\beta}a_{\delta}a_{\gamma}A^{\dagger}_3(\beta_3J'_1J'_2J'_3J)
\end{equation} 
the Wick's theorem and to discard the normal order terms. Within this 
linearization approximation we generate non perturbative solutions of the EoM from the commutator equations of Eq.~(\ref{e4},\ref{e8}), i.e.: 
the eigenvalue equations for the mixed mode system:
\begin{equation}\label{e13}
   \begin{array}{l}
   \displaystyle{ 
   [H,A^{\dagger}_1(\alpha_1J)]|0\rangle
   = \sum_{\beta_{1}} \Omega(2p|2p') A^{\dagger}_1(\beta_1J)|0\rangle } \\
   \displaystyle{
   + \sum_{\beta_2J'_1J'_2} \Omega(2p|3p'1h') 
   A^{\dagger}_2(\beta_2J'_1J'_2J)|0\rangle },
   \end{array}
\end{equation}
and
\begin{equation}\label{e13a}
   \begin{array}{l}
   \displaystyle{
   [H,A^{\dagger}_2(\alpha_2J_1J_2J)]|0\rangle
   = \sum_{\beta_{1}} \Omega(3p1h|2p') A^{\dagger}_1(\beta_1J)|0\rangle } \\
   \displaystyle{
   + \sum_{\beta_{2}J'_1J'_2} \Omega (3p1h|3p'1h') 
     A^{\dagger}_2(\beta_2J'_1J'_2)|0\rangle }.
   \end{array}
\end{equation}
Within the application of the GLA approximation we convert Eqs.~(\ref{e4},
\ref{e8}) in an eigenvalue equation for the configuration mixing wave 
functions (CMWFs) of the model. In fact, the linearization provides the 
additional matrix elements necessary to write the following identity:
\begin{equation}\label{e14}
   \Omega(3p1h|3p'1h') = \langle j_1j_2j_3j_4|v(r)|j'_1j'_2j'_3j'_4\rangle,
\end{equation}
and to introduce the off-diagonal matrix elements which couple the (2p) to 
the (3p1h) subspaces. Now, by writing Eqs.~(\ref{e13},\ref{e13a}) in the 
following matrix form:
\begin{equation}\label{e14a}
   \begin{array}{l}
   \left ( \begin{array}{c}
   \displaystyle{
   [H,A^{\dagger}_1(\alpha_1J)]|0\rangle }   \\
   \displaystyle{
   [H,A^{\dagger}_2(\alpha_2J_1J_2J)]|0\rangle} \end{array} \right ) \\
   \displaystyle{
   = \left ( \begin{array}{cc}
   E_{2p}+\Omega(2p|2p')&\Omega(2p|3p'1h')  \\
   \Omega(3p1h|2p')&E_{3p1h}+\Omega(3p1h|3p'1h')\end{array} \right )
   \left ( \begin{array}{c}
           A^{\dagger}_1(\beta_1J)|0\rangle\\
           A^{\dagger}_2(\beta_2J_1J_2J)|0\rangle \end{array} \right ) } ,
   \end{array}
\end{equation}
and by multiplying to the left with:
\begin{equation}\label{e14b}
   \begin{array}{l}
   \left ( \begin{array}{c}
   \langle0|A_1(\alpha_1J)  \\
   \langle0|A_2(\alpha_2J_1J_2J) \end{array} \right )
   \end{array}
\end{equation}
we generate the eigenvalue equation for the dressed particles:
\begin{equation}\label{e15}
   \begin{array}{l}
   \displaystyle{\sum_{\beta_1\beta_2J'_1J'_2}
   \left ( \begin{array}{cc}
            E_{2p}+\langle A_1(\alpha_1J)|v(r)|A^{\dagger}_1(\beta_1J)\rangle &
	    \langle A_1(\alpha_1J) |v(r)| A^{\dagger}_2(\beta_2J'_1J'_2J)
	    \rangle                                                          \\
            \langle A_2(\alpha_2J_1J_2J)|v(r)|A^{\dagger}_1(\beta_1J)\rangle  & 
	    E_{3p1h}+ \langle A_2(\alpha_2J_1J_2J) |v(r)| 
	    A^{\dagger}_2(\beta_2J'_1J'_2J)\rangle 
	   \end{array} \right)}  \\
           \cdot 
   \left ( \begin{array}{c}  
            \chi_1(\beta_1J) \\
            \chi_2(\beta_2J'_1J'_2J)
	   \end{array}\right)
   = E \left (
           \begin{array}{c} 
            \chi_1(\alpha_1J)\\
            \chi_2(\alpha_2J_1J_2J)\end{array}\right )|0\rangle.
   \end{array}
\end{equation}
In Eq.(\ref{e15}) $E_{2p}=\epsilon^{HF}_{j_1}+\epsilon^{HF}_{j_2}$ and
$E_{3p1h}=\epsilon^{HF}_{j_1}+\epsilon^{HF}_{j_2}+\epsilon^{HF}_{j_3}
 -\epsilon^{HF}_{j_4}$  are the Hartre-Fock energies 
while the $\chi$'s are the projections of the model states:
\begin{equation}  \label{e15a}
   |\Phi^J_{2p}\rangle
   = \chi_1(\alpha_1J) A_1(\alpha_1J)|0\rangle+\chi_2(\alpha_2J_1J_2J)
                       A_2(\alpha_2J_1J_2J)|0\rangle
\end{equation}
to the basic vectors 2p,~3p1h. To conclude, although the (4p-2h) CMWFs are 
not active part of  the model space, they are important for structure 
calculations. One may therefore associate the GLA approximation to a parameter 
which describes the degree of complexity of the model CMWFs. Within the first 
order linearization we obtain the EoM for the shell model while within the 
second and third order linearization approximations we derive the EoM of 
valence particles coexisting with the complex particle-hole structure of the 
excited states.

In this paper we solve Eq.~(\ref{e15}) self-consistently. The solutions for 
the first iteration step are obtained by diagonalizing the eigenvalue 
equation~(\ref{e15}). The first step of the iterative method generates the 
dynamic amplitudes for the two dressed particles, i.e. two particles 
coexisting with the 3p1h structures. With the calculated eigenvectors we 
recompute then the matrix elements $\langle j_1j_2|v(r)|j_1j_2j_3j_4\rangle$ 
and $\langle j_1j_2j_3j_4|v(r)|j'_1j'_2j'_3j'_4\rangle$
and we diagonalize again the eigenvalue equation. The iterations are repeated 
until the stabilization of the energies has been reached. Before performing 
the diagonalization of relative Hamilton's operator in the CMWFs base we have 
to eliminate the spurious center of mass components. In the BDCM this is 
performed, following the calculations of Refs.~[26,~27], by calculating the 
percent weights of spurious states in the model wave functions. These can be 
obtained by evaluating the energy of the center of mass according to the 
following equation:
\begin{equation}\label{c1}
   \begin{array}{l}
   \displaystyle{
   E_R = \int dR\Psi^{\dagger dressed}(j_ij_jJ)(R^2)\Psi^{dressed}(j_ij_jJ)} \\
   \displaystyle{+2\sum_{ij} 
   \int d\vec{r_i}d\vec{r_j}
   \Psi^{\dagger dressed}(j_ij_jJ)(\vec{r_i}\cdot\vec{r_j})
   \Psi^{dressed}(j'_ij'_jJ)}.
\end{array}
\end{equation}
In Eq.~(\ref{c1}) the calculation of the integrals can be performed by 
using the CFT expansion for the (3p1h) states and by considering that for two 
particle states we have:
\begin{equation} 
   \begin{array}{l}
   \langle j_ij_jJ|(\vec{r_i}\cdot\vec{r_j})|j_ij_jJ\rangle\\
   = \frac{4\pi}{3}[\hat{j_i} \hat{j_j}]
   \left ( \begin{array} {ccc}
            j_i & 1 &j_j\\
            -\frac{1}{2}& 0 &\frac{1}{2} 
	   \end{array} 
   \right )2
   \left \{ \begin{array} {ccc}
             j_i & j_j &J\\
             j_i & j_j &1 
	    \end{array} 
   \right \}
   \langle l_i|r|l_j\rangle2,
   \end{array}
\end{equation}
where:
\begin{equation}
   \hat{j} = (2j+1).
\end{equation}
By diagonalizing the above operator in the model space we obtain the energy 
of the center of mass. The overlap with the model space give the degree of 
``spuriosity'' of the different components. The model space which characterize 
the BDCM is formed by adding even particle to a closed-shell nucleus.
The closed shell configuration can be described by a single Slater 
determinant and one can use the Hartree-Fock's theory to obtain the binding 
energy and the single-particle energies. Alternatively one can remark that 
for a closed shell nucleus (Z,N) the single particle energies for the states 
above the Fermi surface are related to the binding energies differences:
\begin{equation}
   \epsilon^>_p = BE(Z,N)-BE^*(Z+1,N),
\end{equation}
and
\begin{equation}
   \epsilon^>_n = BE(Z,N)-BE^*(Z,N+1).
\end{equation}
The single particle energies for the states below the Fermi surface are 
given by:
\begin{equation}
   \epsilon^<_p = BE^*(Z-1,N)-BE(Z,N),
\end{equation}
and
\begin{equation}
   \epsilon^<_n = BE^*(Z,N-1)-BE(Z,N).
\end{equation}
The BE are ground states binding energies which are taken as positive 
values, and  $\epsilon$ will be negative for bound states. $(BE^*=BE-E_x)$ is the 
ground state binding energy minus the excitation energy of the excited states 
associated with the single particle states. Within this method, which recently 
has been reintroduced by B.A. Brown~\cite{bro01}, we derive the single 
particle energies from the known spectra of neighbor nuclei (see Table~(1)).

\begin{table}[tbp]\label{b-I}
\caption{Single-particle scheme and single particle energies (MeV) used to 
form the model CMWFs for the A=6 isotopes}
\begin{tabular}{lcccccccc}
\hline
hole          & $1s_{1/2}$ &       & & & & & &   \\ 
energy        & -20.58     &       & & & & & &   \\ \hline
hole/particle & $1p_{3/2}$ &       & & & & & &   \\ 
energy        & 1.43       &       & & & & & &   \\ \hline
particle      & $1p_{1/2}$ &       $1d_{5/2}$ & $2s_{1/2}$ & $1d_{3/2}$
& $1f_{7/2}$  & $2p_{3/2}$ &       $1f_{5/2}$ & $2p_{1/2}$ \\ \hline
energy        & 1.73       & 17.21 & 22.23 & 23.69
              & 25.23      & 27.18 & 28.33 & 29.67 \\  \hline       
\end{tabular}
\label{level-scheme}
\end{table}

The generalization of the previously defined formalism needed to calculate 
the charge radii of the A=7 (one particle DCM), A=8 (four particles BDCM) 
and A=11 (three particles DCM) is not given explicitly in this paper, 
but will be presented shortly.

\subsection{Results}
\label{sec:2}

In order to perform structure calculations, we have to define a single 
particle base with the relative single-particle energies and to choose the
nuclear two-body interactions. The single-particle energies of these levels 
are taken from the known experimental level spectra of the neighboring nuclei 
and given in Table~(1). For the experimentally unknown single particle energies 
of the fp shells we use the corresponding energies for the mass A=9 nuclei 
scaled accordingly the different binding energies. In this paper we perform 
as in~[12,13] calculations by assuming all levels as bound for the 
particle-particle interaction, we use the G-matrix obtained from Yale 
potential~\cite{sha67}. These matrix elements are evaluated by applying the
$e^S$ correlation operator, truncated at the second order term of the 
expansion, to the harmonic oscillator base with size parameter b=1.76 fm.   
As also explained in Ref.~[4,5] the potential used by the BDCM is separated in 
low and high momentum components. Therefore, the effective model matrix 
elements calculated within the present separation method and those calculated 
by Kuo~[30-33] are pretty similar. The separation method generates 
matrix elements, which are almost independent from the radial shape of the 
different potentials generally used in structure calculations. \\
The particle-hole matrix elements could be calculated from the 
particle-particle matrix elements via a re-coupling transformation. We prefer 
to use the phenomenological potential of Ref.~[34]. The same size parameter 
as for the particle-particle matrix elements has been used.
In Table~(2) the calculated charge radii of $^6$He are compared with the 
radii calculated by the other theoretical models and with the radius obtained 
by the IS theory.
In the theoretical models quoted in this table the calculations of the 
charge distributions and of the charge radii are performed in terms of non-
correlated operators. The correlations are included only in the 
derivation of the $S_2$ effective Hamiltonian.

For the stable $^6$Li the calculated charge radius is equal to 2.55 fm, a value that  
reproduces well the charge radius of 2.55 fm obtained in Ref.~[23] from the 
electron scattering experiments. 
The charge radii given 
in Table~(3) for other lithium isotopes are however larger then those calculated 
in the other quoted theoretical models and then those obtained within the IS theory.
Here also the main difference between the results obtained in the
DCM and BDCM models and those of the other theoretical calculations
has to be found in the treatment of the correlation operator.
In~[9,10] the charge radii are calculated in the ``no core shell model''
which is based on exact solutions of the two particles Schr\"odinger's equation
by considering large computational spaces. The calculations do not include however 
the $S_3$ correlations.
[12-13] presents charge radii evaluated within an accurate Quantum Monte Carlo Method.
The Hamiltonian used includes two- and three-bodies forces in a two-body 
correlated mechanics.
The calculation method of [36-38] is based on a microscopic cluster method in which 
few particles are interacting with the rest nucleus considered in its ground state.
In our model the excitations of the core are associated to the $S_3$ correlation operator
 which increases the charge radius of the Lithium isotopes.
\begin{table}[tbp]\label{b-II}
\label{Radius6He}
\caption{Calculated charge radii for $^6$He in fm compared with the 
results obtained in other theoretical models and with the radius derived 
within the IS theory.}
\begin{tabular}{lc}\hline
charge radius of $^6$He & Model                         \\ \hline
 1.944~\cite{nav98}    & no-core shell model            \\
 2.09~\cite{pip01}     & quantum Monte Carlo technique  \\
 2.25~this work        & $\mathrm{BDCM}$                \\
 2.39~this work        & $\mathrm{BDCM}$ without the folded diagrams        \\
 2.06~this work        & $\mathrm{two~correlated}~1s_{\frac{1}{2}}$-protons \\
 1.99~\cite{wur97}     & Cluster \\   
 1.99~\cite{fun94}     & Cluster \\
 1.99~\cite{var94}     & Cluster \\
 2.054 $\pm$ .014\cite{wan04} & Isotopic Shift~(Exp.)   \\  
\hline
\end{tabular}
\end{table}

\section{Conclusions and Outlook}
\label{conc}

In this contribution we have investigated the effect of the microscopic 
correlation operators on the charge distributions of $^6$He and of the 
Lithium isotopes. The microscopic correlation has been separated in short- 
and long-range correlations according the definition of Shakin~[2,3].
The short-range correlation has been used to define the effective Hamiltonian
of the model while the long-range is used to calculate the structures and the 
distributions of exotic nuclei. As given in the work of Shakin, only the 
two-body short-range correlation need to be considered in order to derive 
the effective Hamiltonian especially if the correlation is of very short 
range. For the long range correlation operator the three body 
component is important and should not be neglected. Within the three body 
correlation operator one introduces in the theory a three body interaction 
which compensates for the use of the genuine three body interaction
of the no-core shell model.

By using generalized linearization approximations and cluster factorization
coefficients we can perform expedite and exact calculations.

Within the calculated correlated distributions we obtain charge radii slightly 
larger than those calculated for non correlated distributions and 
derived by the IS experiments.

The application of the DCM~\cite{tom05} to the two and three electron 
energies and distributions of the Helium and Lithium atoms respectively 
could serve as future motivation for a reevaluation of the IS theory. 
From one side, the calculation of CM of the different isotopes, could 
help to obtain a non-perturbative formulation of the isotopic change of the 
electron transition energies. From the other side the field shift theory could 
include the correct isotopic variation. Since the derivation of the charge 
radii from the two photon experiments is influenced by the precision of the 
theoretical calculations, this new proposed method could contribute 
to evaluate with better precision the charge radii of exotic nuclei.

\begin{table}[tbp]\label{b-III}
\caption{Calculated charge radii for the Lithium isotopes in fm compared 
with the results obtained in other theoretical models and with the radius 
derived within the IS theory (Exp.).}
\begin{tabular}{lcccccc}\hline
Lithium & Exp.(GSI)~\cite{ewa05} & Exp.+Theo.~\cite{tan88} 
        & Theo.~\cite{nav98} & 
Theo.~\cite{pip01} & Theo.~\cite{suz02} & DCM+BDCM\\ \hline
            & rms & rms & rms & rms & rms \\ \hline
  $^6$Li    & 2.51 & 2.47 & 2.22 & 2.54 & -    & 2.55 \\
  $^7$Li    & 2.39 & 2.43 & 2.13 & 2.41 & 2.43 & 2.41 \\
  $^8$Li    & 2.30 & 2.42 & 2.13 & 2.26 & 2.34 & 2.40 \\
  $^9$Li    & 2.22 & 2.34 & 2.16 & 2.21 & 2.27 & 2.42 \\
 $^{11}$Li  & 2.47 & 3.01 & -    & -    & 2.57 & 2.67 \\\hline
\end{tabular}
\end{table}

%\begin{acknowledgements}
%\end{acknowledgements}

%
%
%
%
%
%

\end{document}